\documentclass[a4paper]{jpconf}
\usepackage{graphicx}
\usepackage[tight]{units} %tight is the default
\usepackage{bigstrut}
\usepackage{multirow}

\begin{document}
\title{MINOS Search for Sterile Neutrinos}

\author{Alexandre B. Sousa\footnote{On behalf of the MINOS Collaboration.}}

\address{Department of Physics, Harvard University, Cambridge, Massachusetts 02138, USA}

\ead{asousa@physics.harvard.edu}

\begin{abstract}
Using a NuMI beam exposure of $7.1\times10^{20}$ protons-on-target, the MINOS long-baseline experiment has performed a search for active to sterile neutrino mixing over a distance of 735~km.  Details of the analysis are provided, along with results from comparisons with standard three neutrino oscillations and fits to a 3+1 model including oscillations into one sterile neutrino. An outlook on the future sterile neutrino related contributions from MINOS and the proposed MINOS+ project is also presented.
\\ \\
\noindent {\it Invited paper to NUFACT 11, XIIIth International Workshop on Neutrino Factories, Super beams and Beta beams, 1-6 August 2011, CERN and University of Geneva.\\
\noindent(Submitted to IOP conference series)}
\end{abstract}

\section{Introduction}
The possible existence of one or more light sterile neutrinos, in addition to the three active flavors, has been put forward as an explanation for the observation of an excess appearance of $\bar{\nu}_e$ in a $\bar{\nu}_\mu$ beam over a short baseline, as reported by the LSND experiment. Interest in sterile neutrinos has been rekindled with observations from antineutrino running in the MiniBooNE experiment that are consistent with active to sterile neutrino oscillations driven by at least one additional large mass-square difference $\mathcal{O}$(\unit[1]{eV$^{2}$})\cite{ref:MiniBoone}.  MINOS can probe active to sterile neutrino mixing by measuring the rate of  neutral-current (NC) events at two locations, over a baseline of \unit[735]{km}. Because NC cross-sections are identical among the three active flavors, NC event rates are unaffected by standard neutrino mixing. However, oscillations into a sterile non-interacting neutrino flavor would result in an energy-dependent depletion of NC events at the far site. 

\section{The NuMI Beam and the MINOS Detectors}
MINOS measures neutrinos from the NuMI beam using two detectors: the \unit[980]{ton} (\unit[27]{ton} fiducial) Near Detector (ND), located \unit[1.04]{km} downstream of the beam target at Fermilab; and the \unit[5.4]{kton} (\unit[4.0]{kton} fiducial) Far Detector (FD), placed \unit[735]{km} downstream of the target in the Soudan Underground Laboratory, in Minnesota~\cite{ref:nim}. The energy resolution function for neutrino-induced hadronic showers is approximately $56\%/\sqrt{E}$.  The results presented here were obtained using an exposure of $7.07\times10^{20}$~protons on target taken exclusively with a beam configuration for which the peak neutrino event energy is \unit[3.3]{GeV}. The NuMI beam includes a 1.3\% ($\nu_e+\bar{\nu}_e$) contamination arising from the decay of muons originating in kaon and pion decays.  
\section{Sterile Neutrino Analysis}
In the MINOS detectors, NC interactions give rise to events with a short diffuse hadronic shower and either small or no tracks, whereas charged-current (CC) events typically display a long muon track accompanied by hadronic activity at the event vertex.  The separation between NC and CC events proceeds through selection criteria based on topological variables: events crossing fewer than 47 planes for which no track is reconstructed are selected as NC; events crossing fewer than 47 planes that contain a track are classified as NC only if the track extends less than 6 planes beyond the shower. These selections result in an NC-selected sample with 89\% efficiency and 61\% purity. Highly inelastic $\nu_{_\mu}$~and $\bar{\nu}_\mu$~CC events, where the muon track is not distinguishable from the hadronic shower, are the main source of background for the NC-selected spectrum. Furthermore, the analysis classifies 97\% of $\nu_e$-induced CC events as NC, requiring the possibility of $\nu_e$~appearance to be considered when extracting results.  The predicted NC energy spectrum in the FD is obtained using the ND data. An estimate of the ratio of events in the FD and ND as a function of reconstructed energy, $E_{\rm{reco}}$, is calculated from Monte Carlo simulations. The ratio is multiplied by the observed ND energy spectrum to produce the predicted FD spectrum. To avoid biases, the analysis selections and procedures were determined prior to examining the FD data, following the precepts of a blind analysis. Figures~\ref{fig:NDSpectrum} and~\ref{fig:FDSpectrum} show the reconstructed energy spectra in each detector. 

\section{Results}
\subsection{Comparisons with Three-Flavor Oscillation Scenario}

\begin{figure}[!h]
\centering
\begin{minipage}{17pc}
\centering
\includegraphics[width=0.99\linewidth]{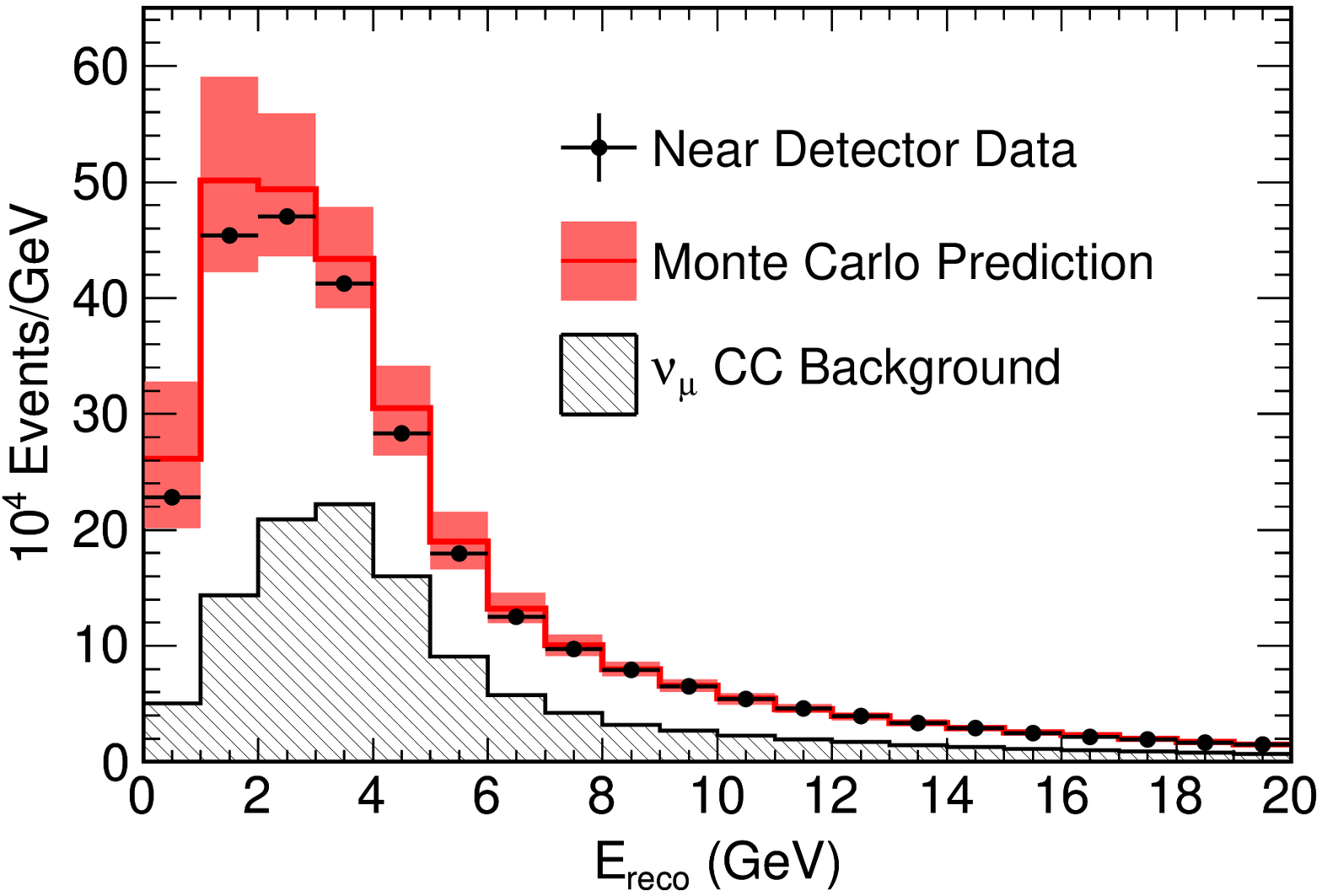}
\caption{\label{fig:NDSpectrum}  Reconstructed energy spectrum of NC-selected events in the ND compared to the Monte Carlo prediction shown with \unit[1]{$\sigma$} systematic errors (shaded band). Also displayed is the simulation of the background from misidentified CC events (hatched histogram).}
\end{minipage}\hspace{2pc}%
\centering
\begin{minipage}{17pc}
\centering
\includegraphics[width=0.99\linewidth]{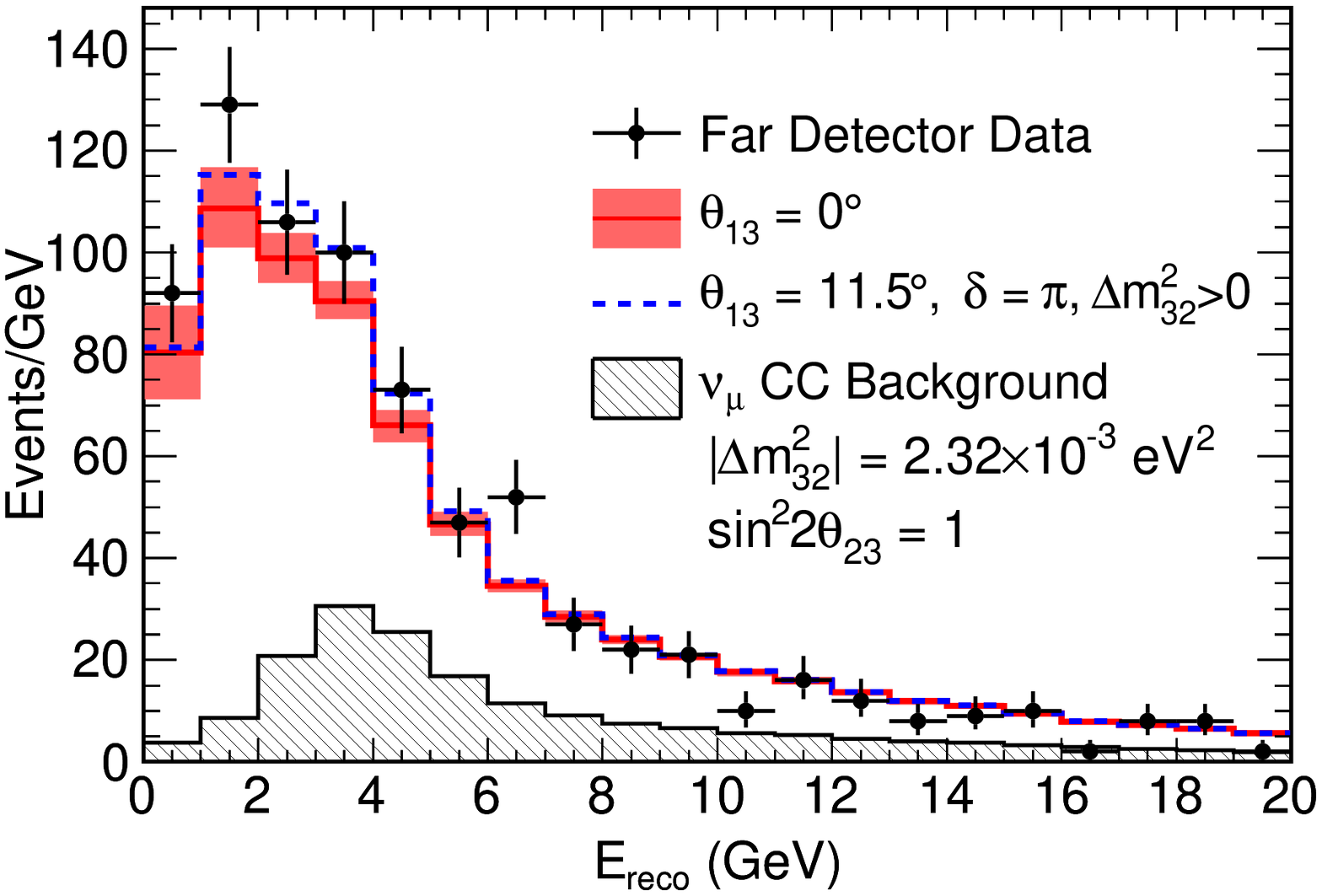}
\caption{\label{fig:FDSpectrum} Reconstructed energy spectrum of NC-selected events in the FD compared with predictions for standard three-flavor mixing with and without $\nu_e$~appearance at the MINOS 90\% C.L. limit~\cite{ref:minosnue}.\\ \\}
\end{minipage} 
\end{figure}

The selection procedures identify 802 NC interaction candidates in the FD, with $754\pm28\rm{(stat)}\pm{37}\rm{(syst)}$ events expected from standard three-flavor mixing (assuming $\theta_{13}=0^\circ$)~\cite{ref:ncprl}. The NC-like reconstructed energy spectrum is shown on Fig.~\ref{fig:FDSpectrum}.
The agreement between the observed and predicted NC spectra is quantified using the statistic $R\equiv\frac{N_{\rm{data}}-B_{\rm{CC}}}{S_{\rm{NC}}}$, where $N_{\rm{data}}$ is the observed number of events, $B_{\rm{CC}}$ is the predicted CC background from all flavors, and  $S_{\rm{NC}}$ is the expected number of NC interactions. The values of $N_{\rm{data}}$, $S_{\rm{NC}}$ and contributions to $B_{\rm{CC}}$ for various reconstructed energy ranges are shown in Table~\ref{tab:nums}. 
\begin{table}[h!]
\caption{\label{tab:nums} Values of the $R$ statistic and its components for several reconstructed energy ranges.  The numbers shown in parentheses include $\nu_e$~appearance with $\theta_{13}=11.5^\circ$ and $\delta_{CP}=\pi$.  The displayed uncertainties are statistical, systematic, and the uncertainty associated with $\nu_e$~appearance.}
\vskip 0.1 cm 
\begin{center}
\lineup
\begin{tabular}{llcccc}
\hline
$E_{\rm{reco}}$ (GeV)      & $N_{\rm{Data}}$ & ~~$S_{\rm{NC}}$~~ & ~~$B^{\nu_\mu}_{\rm{CC}}$~~ & ~~$B^{\nu_\tau}_{\rm{CC}}$~~ & ~~$B^{\nu_e}_{\rm{CC}}$~~ \bigstrut \\ \hline
$0-3$    & 327 & 248.4 & 33.2 & 3.2 & 3.1~(21.5) \bigstrut  \\
$3-120$  & 475 & 269.6 & 156.0 & 9.2 & 31.2~(53.8) \bigstrut \\
\hline
$0-3$ &\multicolumn{5}{l}{$R=1.16\pm0.07\pm0.08-0.08(\nu_e)$} \bigstrut\\
$3-120$ &\multicolumn{5}{l}{$R=1.02\pm0.08\pm0.06-0.08(\nu_e)$} \bigstrut\\
$0-120$ &\multicolumn{5}{l}{$R=1.09\pm0.06\pm0.05-0.08(\nu_e)$} \bigstrut \\
\hline
\end{tabular}\\
\end{center}
\end{table}
The values of $R$ for each energy range show no evidence of a depletion in the NC rate at the FD, supporting the hypothesis that standard three-flavor oscillations explain the data. 

\subsection{Oscillation Fits with a 3+1 Sterile Neutrino Model}
\begin{figure}[h!]
\centering
\includegraphics[width=16pc]{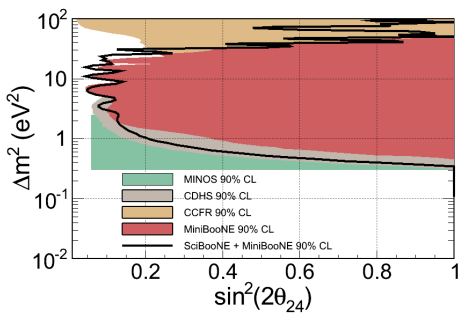}
\hspace{2pc}%
\begin{minipage}[b]{18pc}
\caption{\label{fig:theta24} MINOS exclusion compared to MiniBooNE, CDHS, CCFR $\nu_\mu$ disappearance results. The MINOS 90\%CL excluded region is shown in green. Following the discussion in the text, the bound is assumed to be valid for $0.3~<~\Delta m_{43}^2~<~\unit[2.5]{eV^2}$.  }
\centering
\end{minipage}
\end{figure}

The data are compared with a neutrino oscillation model that allows admixture with one sterile neutrino. In this model, an additional mass scale $\Delta m_{43}^2$ with magnitude $\mathcal{O}$(\unit[1]{eV$^{2}$}) is introduced, along with the assumption that no oscillation-induced change of the neutrino event rate is measurable at the ND site, but rapid oscillations are predicted at the FD location. Based on the magnitude of the systematic uncertainties at the ND, this approximation is assumed to be valid for $0.3<\Delta m_{43}^2<\unit[2.5]{eV^2}$. A detailed description of this model is provided in Ref.~\cite{ref:NCPUBS} and references therein. 
Both the NC-selected energy spectrum shown in Fig.~\ref{fig:FDSpectrum} and the CC-selected spectrum in the FD data are used in the fits to the oscillation models.  Limits on the sterile mixing angles  $\theta_{34} < 26^\circ\,(37^\circ)$ and $\theta_{24} < 7^\circ\,(8^\circ)$ are obtained at the 90\% C.L..  The numbers in parentheses represent the limits extracted for maximal $\nu_e$~appearance. The latter result is presently the most stringent constraint on $\theta_{24}$ for $\Delta m_{43}^2\sim\unit[1]{eV^2}$. A comparison of the MINOS result with other disappearance results is shown in Fig.~\ref{fig:theta24}.
 The coupling between active and sterile neutrinos may also be quantified in terms of the fraction of disappearing $\nu_\mu$~that oscillate into $\nu_s$, $f_{s}~\equiv~P_{\nu_\mu\rightarrow\nu_s}/(1-P_{\nu_\mu\rightarrow\nu_\mu})$, where the $P_{\nu_\mu\rightarrow\nu_x}$ refer to neutrino oscillation probabilities. MINOS places the most stringent constraint to date on this quantity, $f_{s} < 0.22\,(0.40)$ at the 90\%~C.L., where the number in parentheses denotes the limit assuming $\nu_e$~appearance.

A recent paper by Hernandez and Smirnov~\cite{ref:Smirnov} analyzed the MINOS sterile neutrino search results in detail and points out that for large enough values of $\Delta m_{43}$, oscillations at the ND, even if not visible due to the size of systematic uncertainties on the ND energy spectrum, could affect the FD predicted spectrum and reduce the strength of the constraint on $\theta_{24}$ for $\Delta m_{43}^2>\unit[1]{eV^2}$. Moreover, in the same paper and elsewhere in these proceedings~\cite{ref:SmirnovProc, ref:GiuntiProc}, it is pointed out that the combination of the MINOS bound on $\theta_{24}$ with the Bugey bound on $\theta_{14}$ largely rules out the LSND signal region for $0.3<\Delta m_{43}^2<\unit[1]{eV^2}$. Motivated by these remarks, MINOS is working on the inclusion of ND oscillations into the sterile mixing models and will update the analysis with reviewed results in the near future.

\section{Outlook}
\begin{figure}[h!]
\centering
\begin{minipage}{17pc}
\centering
\includegraphics[width=0.99\linewidth]{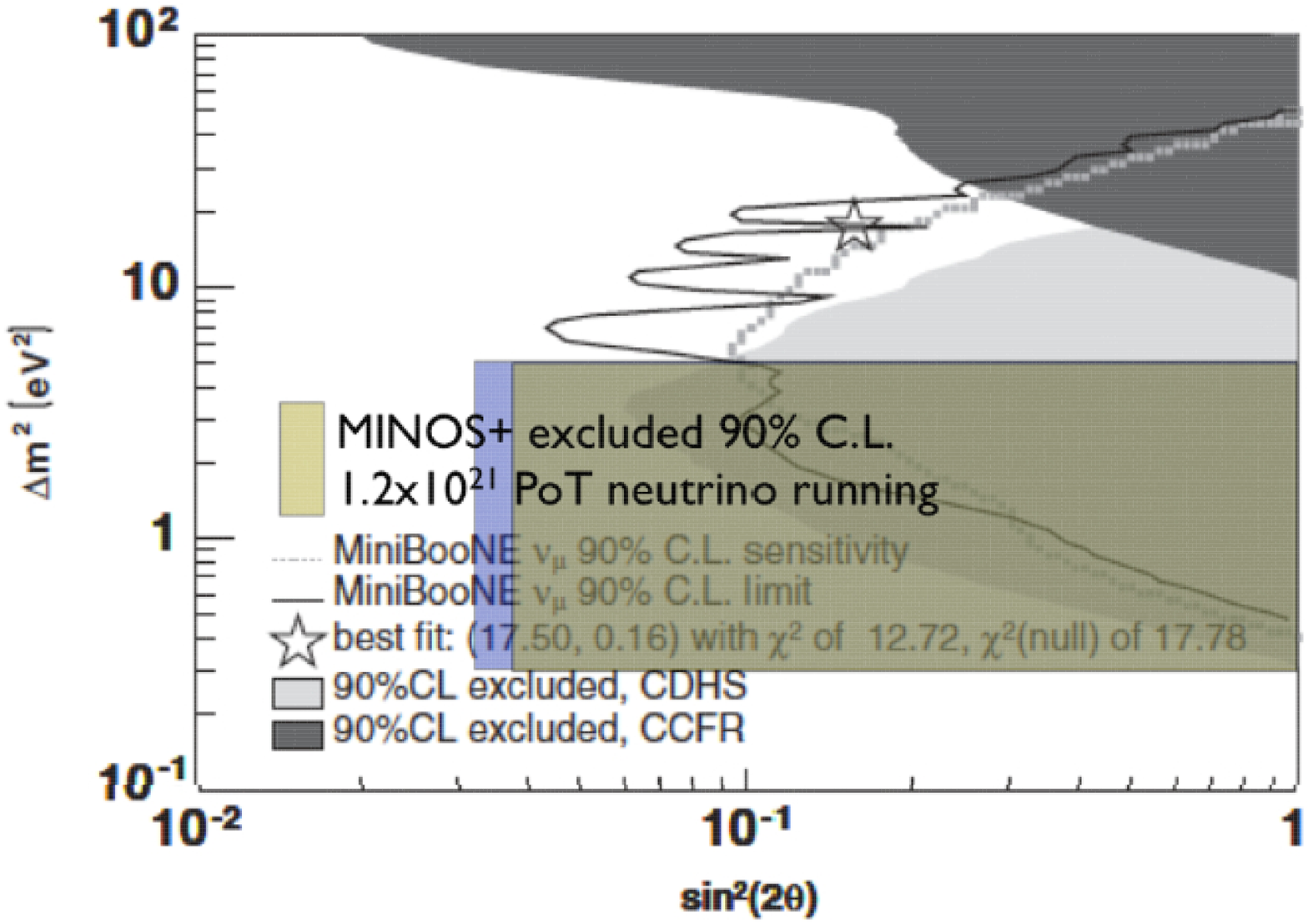}
\caption{\label{fig:minosplusneut} Expected MINOS+ 90\% C.L. exclusion regions for the $\theta_{24}$ mixing angle compared to other disappearance measurements for neutrino running.}
\end{minipage}\hspace{2pc}%
\centering
\begin{minipage}{17pc}
\centering
\includegraphics[width=0.92\linewidth]{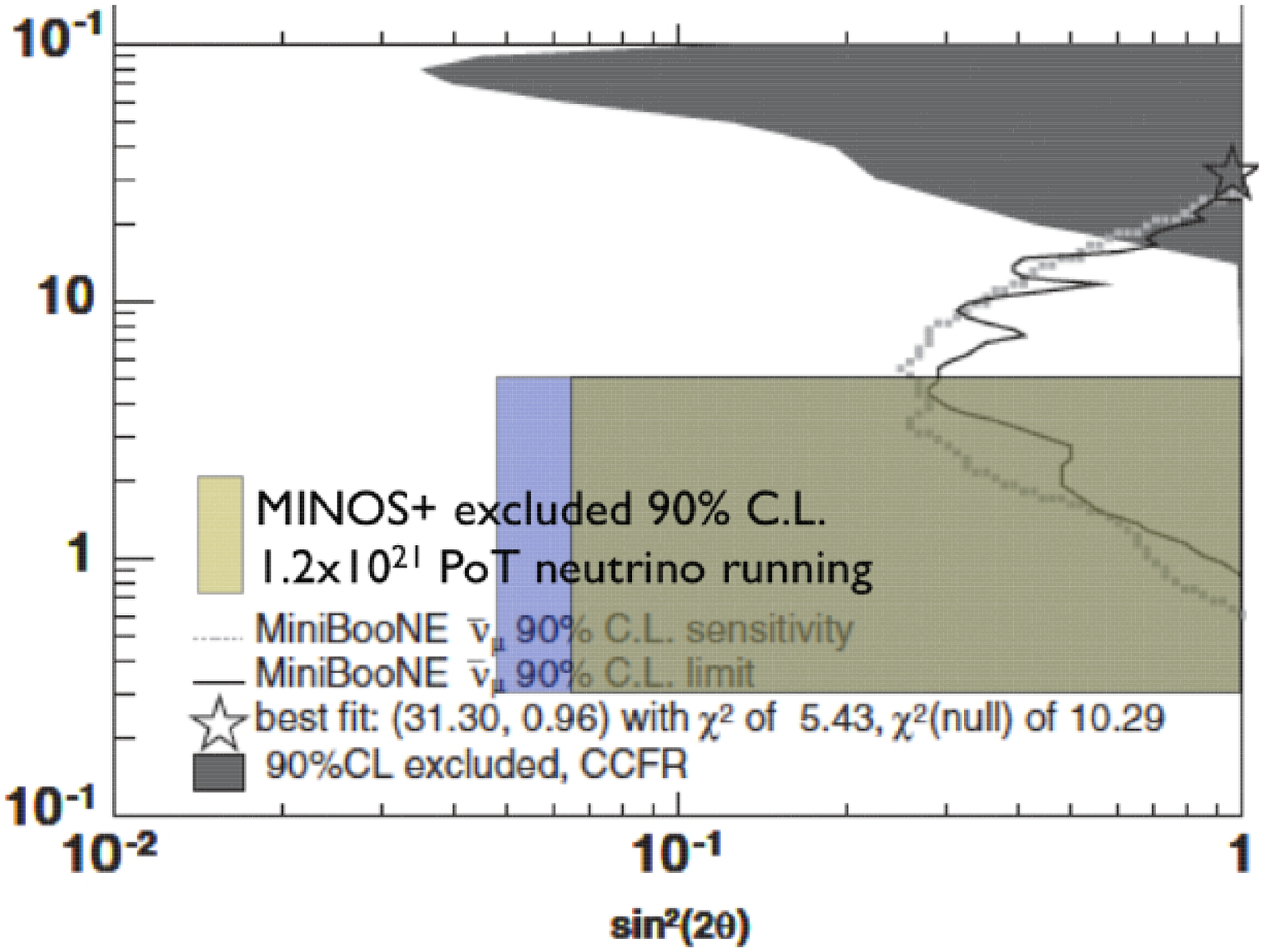}
\caption{\label{fig:minosplusantineut} Expected MINOS+ 90\% C.L. exclusion regions for the $\theta_{24}$ mixing angle compared to other disappearance measurements for antineutrino running.}
\end{minipage} 
\end{figure}

The valuable contributions that MINOS is making to the body of knowledge on sterile neutrino searches will be continued and improved upon by the proposed MINOS+ project. In MINOS+, the MINOS detectors will continue operation during the NO$\nu$A era (starting in 2013), using the NuMI beam upgraded to 700~kW of beam power. MINOS+ will run for three~years in neutrino and three~years in antineutrino mode. The expected sensitivities to $\theta_{24}$ are shown in Figs.~\ref{fig:minosplusneut} and~\ref{fig:minosplusantineut} for neutrino and antineutrino running. MINOS+ can improve the MINOS limits by a a factor of two and place a stringent limit on antineutrino disappearance over short baselines for $\Delta m_{43}^2\sim\unit[1]{eV^2}$. 
Before the MINOS+ era, MINOS also plans to run the sterile neutrino analysis on the $3.3\times10^{20}$~POT of antineutrino data already collected. The main challenges for this analysis arise from limited statistics and feed-down NC interactions from the higher energy neutrino component, which accounts for 58\% of the beam composition when the NuMI horns focus $\pi^-$ and $K^-$. Results of the analysis are expected during 2012.

\end{document}